\documentclass[twocolumn,aps,prl,superscriptaddress,groupedaddress]{revtex4}

\usepackage{graphicx}
\usepackage{dcolumn}
\usepackage{bm}
\usepackage{amsmath,amssymb}

\begin{document}


\title{Enhanced shot noise of multiple Andreev reflections in a carbon nanotube quantum dot in $\mbox{\boldmath $SU(2)$}$ and $\mbox{\boldmath $SU(4)$}$ Kondo regimes}

\author{Tokuro Hata}
\affiliation{Graduate School of Science, Osaka University, Toyonaka, Osaka 560-0043, Japan.}
\author{Rapha\"elle Delagrange}
\affiliation{Laboratoire de Physique des Solides, CNRS, Univ. Paris-sud, Universit\'e Paris, Saclay, 91405 Orsay Cedex, France.}
\author{Tomonori Arakawa}
\affiliation{Graduate School of Science, Osaka University, Toyonaka, Osaka 560-0043, Japan.}
\affiliation{Center for Spintronics Research Network, Osaka University, Toyonaka, Osaka 560-8531, Japan.}
\author{Sanghyun Lee}
\affiliation{Graduate School of Science, Osaka University, Toyonaka, Osaka 560-0043, Japan.}
\author{Richard Deblock}
\affiliation{Laboratoire de Physique des Solides, CNRS, Univ. Paris-sud, Universit\'e Paris, Saclay, 91405 Orsay Cedex, France.}
\author{H\'el\`ene Bouchiat}
\affiliation{Laboratoire de Physique des Solides, CNRS, Univ. Paris-sud, Universit\'e Paris, Saclay, 91405 Orsay Cedex, France.}
\author{Kensuke Kobayashi}
\affiliation{Graduate School of Science, Osaka University, Toyonaka, Osaka 560-0043, Japan.}
\affiliation{Center for Spintronics Research Network, Osaka University, Toyonaka, Osaka 560-8531, Japan.}
\author{Meydi Ferrier}
\affiliation{Graduate School of Science, Osaka University, Toyonaka, Osaka 560-0043, Japan.}
\affiliation{Laboratoire de Physique des Solides, CNRS, Univ. Paris-sud, Universit\'e Paris, Saclay, 91405 Orsay Cedex, France.}

\date{\today}

\begin{abstract}
The sensitivity of shot noise to the interplay between Kondo correlations and superconductivity is investigated in a carbon nanotube quantum dot connected to superconducting electrodes.
Depending on the gate voltage, the $SU(2)$ and $SU(4)$ Kondo unitary regimes can be clearly identified.
We observe enhancement of the shot noise via the Fano factor in the superconducting state.
Its divergence at low bias voltage, which is more pronounced in the $SU(4)$ regime than in the $SU(2)$ one, is larger than what is expected from proliferation of multiple Andreev reflections predicted by the existing theories.
Our result suggests that Kondo effect is responsible for this strong enhancement.
\end{abstract}

\maketitle
{\it Introduction.}|Kondo effect and superconductivity, two typical many-body effects, emerge in various fields of physics such as in heavy fermion systems~\cite{SteglichPRL1979}, mesoscopic systems~\cite{GoldhaberGordonNature1998,FranceschiNatNano2010}, and cold atoms~\cite{NishidaPRL2013, RegalPRL2004}.
Kondo effect arises due to formation of a spin singlet between a localized state and free conduction electrons with the characteristic energy, $k_{\rm B}T_{\rm K}$, where $k_{\rm B}$ is the Boltzmann constant and $T_{\rm K}$ is the Kondo temperature.
An $s$--wave superconductor, for its part, is characterized by a macroscopic wave function constituted of Cooper pairs singlet with a binding energy $\Delta$.
The interplay between those two many-body states has been shown to bring new intriguing fundamental physics such as quantum transitions~\cite{Oguri2007}.
Superconductor--quantum dot--superconductor devices (S--QD--S)~\cite{BuitelaarPRL2002, BuitelaarPRL2003, EichlerPRL2007, JorgensenNanoLett2007, EichlerPRB2009, KimPRL2013} and QD--SQUID~\cite{vanDam2006, Maurand2012, DelagrangePRB2016} tuned in the Kondo regime are ideal platforms to study the interplay between the two effects.

At equilibrium, a supercurrent is carried by Andreev bound states (ABS) formed in the QD.
The Kondo effect manifests itself by screening the magnetic moment of the ground state and induces a first order quantum transition as observed via the current phase relation \cite{vanDam2006,JorgensenNanoLett2007,Maurand2012,DelagrangePRB2016} and spectroscopy experiments \cite{KimPRL2013}.
When the junction is biased with a voltage $V$, non-equilibrium transport occurs through multiple Andreev reflections (MAR)~\cite{BuitelaarPRL2002, BuitelaarPRL2003, EichlerPRL2007}, where an injected quasiparticle at energy, $eV$, transfers effectively a charge $me$ through the junction before escaping when the acquired energy, $m\times eV$, is larger than $2\Delta$~\cite{BTK82, CuevasPRL1998, Nagaev2001, Bezuglyi2001}.

Shot noise, non-equilibrium current fluctuation, is a powerful tool to elucidate such transport mechanisms~\cite{Blanter}.
For instance, large effective charges due to the MAR mechanism depicted above have been detected in the noise measurement in S--quantum point contact (QPC)--S~\cite{CronPRL2001}.
Regarding Kondo effect, experiments have demonstrated that the many-body effect creates a pair backscattering, which induces a large effective charge in the non-linear noise~\cite{Zarchin,Yamauchi, MeydiNatPhys}.
Thus, it is appealing to look for a signature of Kondo effect on the effective charge in a superconducting junction.
However, this topic is still poorly addressed \cite{Avishai2003}, in particular from the experimental point of view.

To explore this interplay, a carbon nanotube (CNT) can be used as a QD to investigate how different symmetries of the Kondo states affect the superconductivity.
Indeed, electrons in a CNT possess two different degrees of freedom, spin and orbital angular momenta, whose degeneracy can be tuned by a gate voltage. 
This enables us to study the $SU(4)$ Kondo effect in the four-fold degenerate case as well as the two-fold degenerate conventional $SU(2)$ Kondo effect~\cite{JarilloHerreroNature2005, Delattre, MakarovskiPRB2007, CleuziouPRL2013, KellerNatPhys2014}.
Recently, it was demonstrated that these symmetries yield different effective charge in the noise~\cite{MeydiNatPhys,MeydiPRL}.
It is thus legitimate to expect a signature of symmetry in the noise associated with MAR, which has never been explored.

In this Letter, we experimentally demonstrate that the different Kondo symmetries appear in the conductance and shot noise of an S--QD--S with unpredicted large Fano factors ($F$).
We have identified three distinct Kondo states, $SU(2)$ at odd filling, $SU(4)$ at odd filling, and $SU(4)$ at even filling.
First, we show how the conductance is sensitive to these three different states.
Then, we demonstrate that the observed divergence of shot noise at low bias voltage cannot be explained quantitatively by the non-interacting MAR theory.
This result clearly points out a lack in the theory of shot noise in interacting systems and provides a well-defined quantitative measurement, which should trigger new developments in the many-body physics.

{\it Experimental setup.}|The sample consists of a CNT connected to Pd($6$~nm)/Al($70$~nm) superconducting leads, where $\Delta$ is $45\ {\rm \mu eV}$~\cite{MeydiNatPhys}.
Measurements are carried out in a dilution refrigerator at $20\ \rm{mK}$.
We measure differential conductance ($G(V)\equiv dI/dV$) at source-drain voltage $V$ using a lock-in technique with an ac excitation of $2\ \rm{\mu V}$.
The current noise spectral density $S_{\rm I}$ is measured through a resonant LC circuit ($2.58\ \rm{MHz}$) connected to a low temperature amplifier~\cite{ArakawaAPL2013}.
The device is set either in the normal (N) state or in the superconducting (S) state by switching on and off a small in-plane magnetic field of $0.08\ \rm{T}$, respectively.
Recently, N-state characteristics were studied by conductance and shot noise measurement~\cite{MeydiNatPhys, MeydiPRL}.
Since we focus on the noise at a few MHz, measurement of the DC intrinsic supercurrent is beyond the scope of the present experiment.
Indeed, supercurrent measurement requires a well controlled electromagnetic environment and the use of a few $\rm{kHz}$ low-pass noise filter to avoid external noise which is not compatible with our noise set-up.
However, our setup is well suited for voltage biased conductance and noise measurements on which we focus here.

\begin{figure}[tp]
\center \includegraphics[width=1\linewidth]{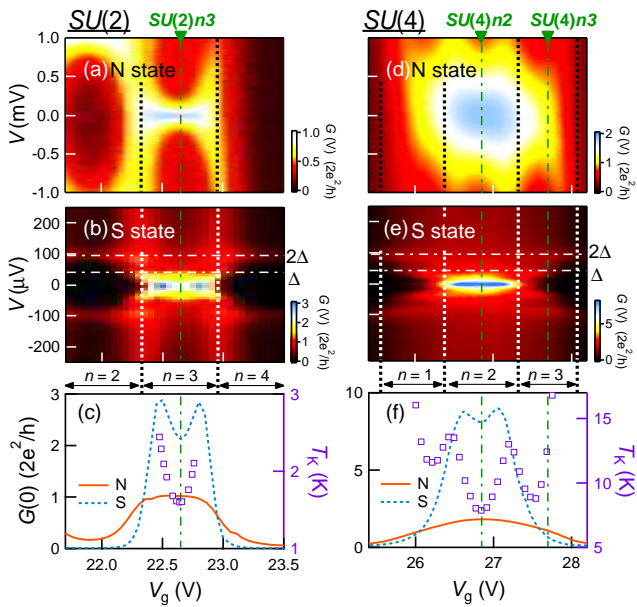}
\caption{(color online) (a)(b) Color plot of $G(V)$ as a function of source-drain voltage ($V$) and gate voltage ($V_{\rm g}$) in the N-state and in the S-state for the $SU(2)$ symmetry, respectively. The value, $\Delta=45\ \rm{\mu eV}$, is extracted from the line at $eV=2\Delta$ which does not depend on $V_{\rm g}$. (c) $V_{\rm g}$ dependence of $G(0)$ in the N-state (solid curve) and S-state (dashed curve) for $SU(2)$. Squares represent Kondo temperature $T_{\rm K}$. $G(0)$ reaches the unitary limit $G=G_Q$ in the N-state of $SU(2)$, ensuring a perfect left/right symmetric coupling of the QD. $n$ represents the number of electrons in the last shell of the CNT dot. (d)(e) Color plot of $G(V)$ as a function of $V$ and $V_{\rm g}$ in the N-state and in the S-state for the $SU(4)$ symmetry, respectively. (f) $V_{\rm g}$ dependence of $G(0)$ for the N-state (solid curve) and S-state (dashed curve) and $T_{\rm K}$ (square marks) for $SU(4)$. This work focuses on the three states, ``$SU(2)n3$'', ``$SU(4)n3$'', and ``$SU(4)n2$'', whose positions are represented on the top of the figures.}
\label{001}
\end{figure}

{\it Kondo effects in the N-state.}|Figures~\ref{001}(a) and (d) show a color plot of  $G(V)$ as a function of $V$ and gate voltage ($V_{\rm g}$) in the  N-state for $SU(2)$ and $SU(4)$, respectively.
The QD filling, which is controlled by $V_{\rm g}$, consists of successive shells of four electrons.
We denote the number of electrons on the last occupied shell by $n=1,\ 2,\ 3$, and $4$.
The $SU(N)$ symmetry can be distinguished from the shape of these plots.
The Kondo effect always yields a maximum in the zero bias conductance $G(0)$. 
It results in horizontal bright regions called Kondo ridges.
For $SU(2)$ (Fig.~\ref{001}(a)), this maximum only appears for odd filling ($n=1$ [not shown here] and $n=3$).
For $SU(4)$ (Fig.~\ref{001}(d)), $G(0)$ takes its maximum for $n=2$ as well as for $n=1$ and $3$, yielding an extended wide Kondo ridge~\cite{MeydiNatPhys,MeydiPRL}, which is a typical behavior of $SU(4)$ Kondo state~\cite{JarilloHerreroNature2005, MakarovskiPRB2007, CleuziouPRL2013, KellerNatPhys2014}. 

$V_{\rm g}$ dependence of $G(0)$ in the $SU(2)$ case is shown as a solid curve in Fig.~\ref{001}(c).
The conductance at $n=3$ reaches the unitary limit, $G_{\rm Q}\equiv 2e^2/h$, corresponding to a single perfectly transmitting channel (transmission probability: $T\simeq1$), as confirmed by the absence of shot noise~\cite{MeydiNatPhys}.
We label this state ``$SU(2)n3$''.
The Kondo temperature $T_{\rm K}$ (square marks in Fig.~\ref{001}(c)) is obtained from the temperature dependence of $G(0)$~\cite{MeydiNatPhys}.
Its minimum value is $1.6\ \rm{K}$ ($k_{\rm B}T_{\rm K}=137\ \rm{\mu eV}$), which is about three times larger than $\Delta$.

The $SU(4)$ conductance shown in Fig.~\ref{001}(f) takes the value $G_{\rm Q}$ at quarter fillings for $n=1$ and $3$, while it almost reaches its maximum $2G_{\rm Q}$ at half filling for $n=2$~\cite{MeydiNatPhys}.
The Kondo state at $n=3$ is labeled ``$SU(4)n3$'', which corresponds to two half transmitting channels ($T_1=T_2=1/2$).
The Kondo state at $n=2$ (labeled as ``$SU(4)n2$'') possesses two perfect channels, $T_1=T_2\simeq 1$. 
Square marks in Fig.~\ref{001}(f) represent $T_{\rm K}$, which is estimated from the width of the conductance because the large $T_{\rm K}$ makes the conductance hardly depend on temperature up to $800\ \rm{mK}$.
Note that the experimental condition is in the regime $k_{\rm B}T_{\rm K}\gg\Delta$.

To compare these three states ($SU(2)n3$, $SU(4)n3$, and $SU(4)n2$), it is significant to have in mind that both $SU(2)n3$ and $SU(4)n2$ correspond to half filling with electron-hole symmetry, yielding perfect transmissions.
In contrast, $SU(4)n3$ is at quarter filling without electron-hole symmetry, and has two half transmitting channels, $T_1=T_2=1/2$.

{\it Kondo effects in the S-state.}|
Figure \ref{001}(b) presents a color plot of $G(V)$ in the S-state for $SU(2)$.
A peak at $eV=2\Delta$ in $G(V)$ due to quasiparticle tunneling above the gap appears at every filling factor ($n$).
The zero bias conductance $G(0)$, represented as a dashed curve in Fig.~\ref{001}(c), is enhanced on the Kondo ridge ($n=3$), whereas it vanishes for $n=2\ $and $4$.
Note that the $G(0)$ has a minimum at half filling like the Kondo temperature, which is consistent with Ref.~\cite{BuitelaarPRL2002}.

The image plot in Fig.~\ref{001}(e) and the dashed curve in Fig.~\ref{001}(f) show $SU(4)$ Kondo effect in the S-state.
The zero bias conductance is only enhanced around $n=2$, whereas it is not the case at quarter filling ($SU(4)n3$).
Given the above discussion for the $SU(2)$ state, this result demonstrates that conductance is enhanced only when electron-hole symmetry holds.
Like in the $SU(2)$ symmetry, the conductance has a local minimum at half filling.
Although a zero resistance state cannot be detected as expected above, it is possible to extract a value of the DC supercurrent ($I_{\rm c}$) from the curve $G(V)$ by taking the circuit impedance into account (see Supplement).
Interestingly, this supercurrent follows the same shape as $G(0)$ and $T_{\rm K}$ as a function of $V_{\rm g}$.

In the following, we compare $G(V)$ and current noise $S_{\rm I}$ in the three Kondo states to show how Kondo symmetry manifests itself in the S-state.

\begin{figure}[tp]
\center \includegraphics[width=1\linewidth]{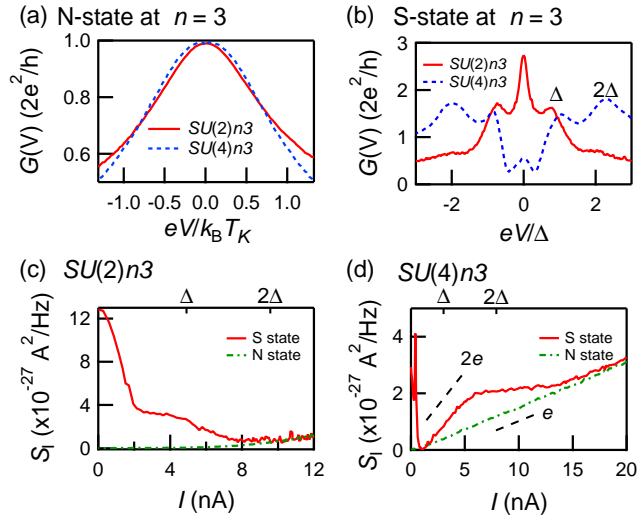}
\caption{(color online) (a) $eV/k_{\rm B}T_{\rm K}$ dependence of $G(V)$ at $n=3$ in the N-states for $SU(2)n3$ (solid curve) and $SU(4)n3$ (dashed curve). The conductance shapes are almost indistinguishable. (b) $eV/\Delta$ dependence of $G(V)$ in the S-state for $SU(2)n3$ (solid curve) and $SU(4)n3$ (dashed curve). For $SU(2)$, a single perfectly transmitting channel  allows high order MAR and high $G(0)$, while two channels with low transmission ($T_{1}=T_{2}=0.5$) suppress $G(V)$ at low bias for $SU(4)n3$. (c) $I$ dependence of $S_{\rm I}$ for $SU(2)n3$ in the N- and S-states (solid and dashed curves, respectively). The enhancement around zero bias is due to high order MAR. (d) $I$ dependence of $S_{\rm I}$ for $SU(4)n3$ in N- and S-states (solid and dashed curves, respectively). The dashed lines correspond to $e^{\ast}=e$ with $F=1-T=1/2$ and $2e$ with $F=1-T^2=3/4$ (see text and Supplement).}
\label{002}
\end{figure}

{\it Conductance at $n=3$.}|We start to compare $SU(2)n3$ and $SU(4)n3$.
Figure~\ref{002}(a) shows $G(V)$ as a function of the rescaled voltage $eV/k_{\rm B}T_{\rm K}$ in the N-state.
The two curves are perfectly superimposed and reach the value $G=G_{\rm Q}$ in the both cases.
However, in the S-state, Fig.~\ref{002}(b) shows a completely different shape for each state.
The zero bias conductance of $SU(2)n3$ is enhanced to $3G_{\rm Q}$, whereas it is suppressed to $0.5G_{\rm Q}$ for $SU(4)n3$.

\begin{figure}
\center \includegraphics[width=85mm]{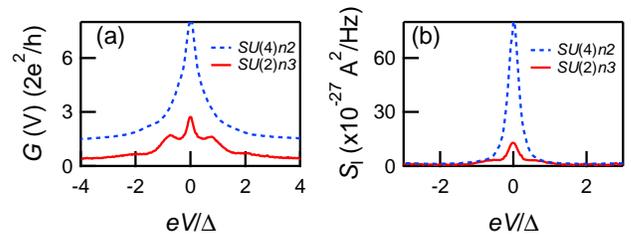}
\caption{(color online) (a) $eV/\Delta$ dependence of $G(V)$ in the S-state for $SU(2)n3$ and $SU(4)n2$, which correspond to the electron-hole symmetry case. Those two states have perfectly transmitting channels leading to high order interfering MAR, which enhances $G(V)$ at low voltage. (b) $eV/\Delta$ dependence of $S_{\rm I}$ in the S-state for $SU(2)n3$ and $SU(4)n2$. Here, interfering MAR enhances the low bias shot noise.}
\label{003}
\end{figure}

In the absence of quantitative prediction for a Kondo-correlated S--QD--S junction, the qualitative difference between the $SU(2)$ and $SU(4)$ Kondo symmetries can be caught by the single particle theory of MAR for QPC~\cite{CuevasPRL1998,BTK82}.
As depicted in the introduction, a quasiparticle injected at energy $eV<2\Delta$ experiences $m-1$ Andreev reflections before escaping at energy $meV\geq2\Delta$ (with $m\geq2$).
Since the probability of MAR transport depends on the transmission as $T^m$, the following picture emerges~\cite{CuevasPRL1998}.
For low transmission, high order processes have very low probabilities, and as a result, conductance vanishes significantly at low voltage.
On the other hand, high order MAR processes are possible for high transmission, yielding a high conductance at low voltage.
It is established that the MAR processes happen sequentially in low transmission junction, which transfer well-defined integer effective charge~\cite{CronPRL2001, CuevasPRL1998}, $e^*/e\equiv m=1+{\rm Int}\left(\dfrac{2\Delta}{eV}\right)$,
whereas interferences between MAR lead to a continuous increase of the effective charge in a high transmission junction when lowering the bias voltage.
In the latter case, transport is equivalently described as Landau-Zener transitions between ABS~\cite{AverinBardas}.

It is thus straightforward to explain qualitatively the shape of $G(V)$ in the S-state shown in Fig.~\ref{002}(b) for the different Kondo symmetries at $n=3$.
$SU(2)n3$ has one single high transmission channel, allowing MAR interferences and enhancing conductance at low voltage. 
In contrast, $SU(4)n3$ has two semi-transparent channels, $T=1/2$, which suppresses the conductance at low voltage in the S-state.
In addition, $\Delta$ and $2\Delta$ peaks in $G(V)$ appear more clearly for $SU(4)n3$ as expected in the sequential regime.
There is a small peak at $V=0$, which is probably due to thermally activated MAR~\cite{RatzNewJofPhys2014} since its amplitude increases with increasing temperature.

{\it Conductance at half filling.}|
Now, we compare Kondo states at half filling ($SU(2)n3$ and $SU(4)n2$).
Similarly to $SU(2)n3$, $SU(4)n2$ has perfect channels and shows high order interfering MAR, which greatly enhance $G(V)$ at low voltage (Fig.~\ref{003}(a)).
More quantitatively, in the Supplement, it is shown that $SU(2)n3$ has a conductance smaller than the non-interacting prediction, whereas the $SU(4)n2$ is better reproduced by the model.
Note that the theory assumes a constant density of states in the normal part of the junction on the scale of the gap.
This is valid for $SU(4)n2$ with $k_{\rm B}T_{\rm K}/\Delta\simeq15$, but not for $SU(2)n3$ with $k_{\rm B}T_{\rm K}/\Delta\simeq3$.
Thus, the difference between the half-filling states may be due not only to the symmetry but also to the energy dependence of the transmission in the N-state.

\begin{figure}[tbp]
\center \includegraphics[width=1\linewidth]{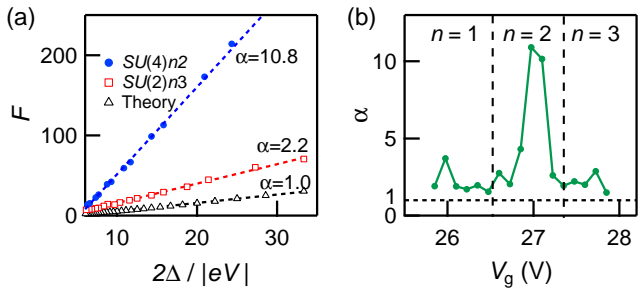}
\caption{(color online) (a) Fano factor in the S-state at half-filling. $1/|V|$ dependence of $F=S_{\rm I}/2e|I|$ in the S-state for $SU(2)n3$ (squares), $SU(4)n2$ (circles), and numerical calculation result for a QPC (triangles). The points are fitted with Eqn. \ref{FanoSuper}. $\alpha$ is the enhancement factor in the Kondo regime compared to an S-weak link. (b) $V_{\rm g}$ dependence of the superconducting noise enhancement $\alpha$ in the $SU(4)$ case. $\alpha$ shows its maximum in the state $SU(4)n2$ at the electron-hole symmetric point.}
\label{004}
\end{figure}

{\it Shot noise.}|
We clarify microscopic transport processes in more detail with shot noise measurement.
First, let us see the difference between $SU(2)n3$ and $SU(4)n3$.
Figures~\ref{002}(c) and (d) display the current dependence of $S_{\rm I}$ for $SU(2)n3$ and $SU(4)n3$ in the N- and S-states, respectively.
In the N-state, shot noise in a non-interacting system is given by $S_{\rm I} = 2eIF$.
We previously confirmed~\cite{MeydiNatPhys} $F \sim 0$ for $SU(2)n3$ and $SU(4)n2$, which correspond to one and two perfect channels, respectively.
In contrast, $F = 1/2$ for $SU(4)n3$ due to two half transmitting channels.

In the S-state, we first note that we recover the same slope as in the N-state for $eV\gg 2\Delta$ for the two states (Fig.~\ref{002}(c) and (d)).
For an odd filling, the two symmetries have different behavior at low current.
Both present broad peaks for the current $I$ corresponding to $eV=\Delta$.
Interestingly, for $SU(4)n3$, we observe that the noise at $\Delta$ decreases with a slope corresponding to $e^{\ast} = 2e$ with a Fano factor of $3/4$, namely, $S_{\rm I} = 2(2e)I \times 3/4$ [see the dashed line in Fig.\ref{002} (d)].
This might be explained by naively assuming that the transmission probability of the first order MAR is given by $T^2 = (1/2)^2$ and thus $F=1-T^2 = 3/4$ by simple analogy with the shot noise in a single-channel QPC (see Supplement).
The sharp peak at $I<1\ \rm{nA}$~ presumably corresponds to thermally activated ABS, which can be observed as the small peak in $G(0)$ (see Fig.~\ref{002}(b)).

In contrast, in the $SU(2)$ symmetry, the noise is greatly enhanced at low current below $\Delta$ (solid curve in Fig.~\ref{002}(c)).
Such an enhancement is also observed in the $SU(4)n2$ state (Fig.~\ref{003}(b)).
It can be quantitatively analyzed via the Fano factor, $F\equiv S_{\rm I}/2eI$.
Fano factor as a function of $1/V$ is shown in Fig.~\ref{004} (a).
This plot is the central result of the present work.
In this representation, theoretical curves, which do not include possible Kondo correlations, for QPC of any transmission fall on the same line corresponding to the expected number of transmitted pairs, $F_{\rm QPC}\simeq 2\Delta/eV$ (triangles in Fig. \ref{004}(a))~\cite{CuevasPRL1998}.
Our experimental results (circles and squares) are also almost linear in $1/V$.
However, the slope is enhanced by a factor $\alpha$.
Thus, we have fitted our result using the formula,
\begin{align}
F= \frac{2\Delta}{e}\times\frac{\alpha}{|V|}.
\label{FanoSuper}
\end{align}
We obtain $\alpha\simeq 2.2$ for $SU(2)n3$ ($k_{\rm B}T_{\rm K}/\Delta\simeq3$) and $\alpha\simeq 10.8$ for $SU(4)n2$ ($k_{\rm B}T_{\rm K}/\Delta\simeq15$).
In the $SU(4)$ symmetry, we have measured the dependence of $\alpha$ with the filling factor (Fig. \ref{004}(b)).
It has clearly a maximum at half filling and decreases around $\alpha=2$ at quarter filling ($SU(4)n3$).
In addition, we have checked that this effect disappears when temperature increases (see Supplement).
These results suggest that MAR interferences are affected by the Kondo effect, which enhances their noise.
This effect is maximum in the limit, $T_{\rm K}\gg\Delta$, and probably in the $SU(4)$ symmetry.
Although it has been shown that at very low voltage, the noise of a S--QD--S junction is very non-linear and depends on microscopic details of the QD such as the relaxation time~\cite{Averin1996,Martin-Rodero1996}, we argue that universality of Kondo effect should make our observation very general.

{\it Conclusion}|
In conclusion, we have measured non-equilibrium superconducting proximity effect through a CNT QD in the $SU(2)$ and $SU(4)$  Kondo regime.
Using the non-interacting MAR theory, the difference in the conductance and the shot noise can be qualitatively explained from the different sets of transmissions given by the different symmetries of the Kondo states. 
However, no quantitative agreement can be achieved, calling for a theory of fluctuations for the Kondo-superconductivity interplay in the non-equilibrium regime.
In particular, we observe a strong enhancement of the Fano factor at low bias which bears the signature of Kondo effect and is more pronounced in the $SU(4)$ symmetry.
These results should trigger further developments on fluctuations in competing many-body systems.

We specially thank J. C. Cuevas for providing us numerical data and for discussions.
We acknowledge valuable discussions with M. Filippone, S. Gu{\'e}ron, A. Levy-Yeyati,  A. Oguri, F. von Oppen, and R. Sakano.
This work was partially supported by JSPS KAKENHI Grant Numbers JP26220711, JP26400319, JP16K17723, JP15K17680, JP15J01518, JP25103003, and JP15H05854, Yazaki Memorial Foundation for Science and Technology, the French Programs ANR MASH (Program No. ANR-12-BS04-0016), DYMESYS (Program No. ANR 2011-IS04-001-01), DIRACFORMAG (Program No. ANR-14-CE32-0003), and JETS (Program No. ANR-16-CE30-0029-01).

\end{document}